\documentclass[nofootinbib,aps,12pt]{revtex4}

\usepackage{amssymb,amstext,amsmath,amsthm}
\usepackage[dvips]{graphicx}
\usepackage{latexsym}
\usepackage{psfrag}
\usepackage{amsfonts}
\usepackage{bbm}
\usepackage{color}

\newcommand{\Ref}[1]{(\ref{#1})}

\newcommand{\eqa}{\begin{eqnarray}}
\newcommand{\neqa}{\end{eqnarray}}
\newcommand{\equ}{\begin{equation}}
\newcommand{\nequ}{\end{equation}}

\def\om{\omega}
\def\w{\wedge}

\def\d{\delta}
\def\f{\frac}

\usepackage{bbm}

\newcommand{\scr}{\rm\scriptscriptstyle}

\newcommand{\SO}{\mathrm{SO}}

\let\eps=\epsilon

\newcommand{\GN}{G_{\scr N}}
\newcommand{\al}{\alpha}
\newcommand{\be}{\beta}\newcommand{\ga}{\gamma}\newcommand{\la}{\lambda}

\begin{document}

\title{\Large\bf A note on the Plebanski action with cosmological constant and an Immirzi parameter}
\author{Lee Smolin${}^{a}$ and Simone Speziale${}^{b}$
\\ \medskip
\small ${}^a$\emph{Perimeter Institute for Theoretical Physics, 31 Caroline St. N, ON N2L 2Y5, Waterloo, Canada} \\ {\small ${}^b$\emph{Centre de Physique Th\'eorique, CNRS-UMR 6207, Case 907 Luminy, Marseille 13288, France}}}
\email{lsmolin@perimeterinstitute.ca, simone.speziale@cpt.univ-mrs.fr}

\date{August 24, 2009}

\begin{abstract}
We study the field equations of the Plebanski action for general relativity when both the cosmological constant and an Immirzi parameter are present. We show that the Lagrange multiplier, which usually gets identified with the Weyl curvature, now acquires a trace part. Some consequences of this for a class of modified gravity theories recently proposed in the literature are briefly discussed.
\end{abstract}

\maketitle

\section{Introduction}

At the heart of the last few decades progress in quantum gravity is the fact that it is possible to formulate
general relativity in terms of simple polynomial equations.  The translation of the Einstein equations into 
operator equations would otherwise be an infinitely hard problem, as it requires defining operator products 
of arbitrary order to represent non-polynomial functionals of the basic operators.  Indeed, the successes
of loop quantum gravity and spin foam models stem from the fact that the Ashtekar and related formulations
are polynomial.   

The simplest non-linear equation is of quadratic order, and hence the simplest action principles for an interaction 
field theory will be cubic in basic variables.  It is interesting that cubic actions can be written for general relativity, such as the ones proposed by Deser \cite{Deser} and Plebanski \cite{Plebanski}. The Plebanski action, which is the object of interest of this paper, is obtained adding suitable auxiliary fields. It was originally written in the 1970's \cite{Plebanski}, and rediscovered by \cite{Capo2} who were seeking an action principle for Ashtekar variables.

In the Plebanksi action, and the Hamiltonian formulations which follow from it, the metric is no longer the configuration variable of the theory.  Instead, the configuration variables is a connection, $A_a$ valued in a Lie algebra $\mathfrak{g}$.
In the original and simplest case, $\mathfrak{g}=\mathfrak{su(2)}$, the left handed chiral subalgebra of the local Lorentz group of general relativity. The action, which we will describe below, is the simplest cubic combination of the curvature of $A_a$ and some simple auxiliary fields, and it leads to a theory equivalent to general relativity. 

Since then the theory has been extended in several ways. First, a class of modified gravity theories is obtained by simply introducing more complicated dependence on the auxiliary fields \cite{Krasnov}.  Secondly, $\mathfrak{g}$ can be extended to the full algebra of Lorentz group, which also gives a formulation of general relativity \cite{Capo2,Mike,De Pietri}.  Third, these two extensions have been combined, yielding a class of modified gravity theories with additional degrees of freedom \cite{Lee, Alexandrov, noi2}.  Finally, a further extension is to take a group which contains a product of the local Lorentz group and an internal gauge group \cite{Lee}.  This yields 
a unification of general relativity with Yang-Mills and Higgs fields \cite{noi1}.  

During the investigation of \cite{Lee,noi1}, a couple of subtle technical issues arose, which concern the role of the Immirzi parameter and its interplay with the cosmological constant, in the various extensions of the Plebanski formulation.  The resolution of these issues is the subject of this short paper.  Hence this is a paper which is likely to be of interest only to specialists immersed in the study of the Plebanski theory and its extensions.  But we hope that the reader, whether specialist or not,  will be amused by the subtle interplay of the  Immirzi and cosmological constants in these extensions of general relativity.  

In the next section we study the non-chiral Plebanski theory, with the full lorentz connection,  with both a cosmological constant and Immirzi parameter.  We focus on the interplay of the two constants in the equations of motion, uncovering and we hope elucidating some subtleties which to our knowledge, have not been described in the literature.

In section 3 we consider the simplest of the extensions of the non-chiral Plebanski theory, which involve higher order terms in the auxiliary fields, first introduced in \cite{Lee}.  With the preparation from section 2 we are able to untangle the interplay of the Immirzi parameter and cosmological constant.  We find a perhaps surprising result, which is that the field equations are only consistent for a few special values of the Immirzi parameter.  Once chosen, we show that deSitter (or AdS) spacetime is a solution to the full equations of motion.  This result is essential to recent studies of the extension which includes
a unification with Yang-Mills and Higgs fields.  Beyond this, the full structure of the equations of motion will
be elucidated in \cite{noi2}, where it is shown that the gravitational sector of the theory is a bimetric theory.  

Throughout the paper, we take units $16\pi \GN=1$, and use greek letters for spacetime indices and latin letters for Lorentz indices.
We take Euclidean signature, but all our formulas can be easily modified to Lorentzian signature. 
$[a,b]$ means normalized antisymmetrization, and we also use the shorthand notation
${\mathbbm 1} = \d^{IJ}_{KL} = \f12 (\d^I_K \d^J_L-\d^I_L \d^J_K)$,
$\d_{IJKL} = \f12 (\d_{IK} \d_{JL}-\d_{IL} \d_{JK})$, $\star=\f12\eps^{IJ}{}_{KL}$.

\section{Plebanski equations}

The non-chiral Plebanski action is
\equ\label{S}
S(B,\om,\phi) = \int B^{IJ} \w F_{IJ}(\om) - \f12 \phi_{IJKL} \, B^{IJ} \w B^{KL} + V(B)
\nequ
where $F(\om)$ is the curvature of an $\mathfrak{so(4)}$ connection $\om$, and $B^{IJ}$ a 2-form with values in the algebra.
The potential $V(B)$ will be related below to the cosmological constant, and extended to $V(\phi,B)$ in the modified theory discussed in Section \ref{SecMod}.
The field $\phi^{IJKL}$ is a Lagrange multiplier,
which upon variation enforces constraints on the $B$ field. The Lagrange multiplier is symmetric under exchange of the first and the second pair, and antisymmetric within each pair. 
In addition, we impose on $\phi$ the constraint
\equ
\eps_{IJKL}\phi^{IJKL}=0. 
\label{epconstraint}
\nequ 
It has therefore the same symmetries as the Riemann tensor and 20 components. 

In this section we will focus attention to the values  
taken by the Lagrange multiplier $\phi$ in solutions of the field equations.  
In the standard case known in the literature, without Immirzi parameter, it becomes the Weyl tensor, and it is correspondingly traceless. 
As we show below, the situation is more subtle when both the Immirzi parameter and a cosmological constant are present. Untangling the equations, we see two new phenomena: 
first, the cosmological constant is a function of the Immirzi parameter;
second, $\phi$ acquires a trace, proportional to the cosmological constant.
Its on-shell value is now the Weyl tensor plus a trace.

Concerning the cosmological constant term, there are two possible ``volume forms'' that one can consider, 
$(B\w B) \equiv \d_{IJKL}B^{IJ} \w B^{KL}$ and $(B\w\star B) \equiv (1/2)\eps_{IJKL} B^{IJ} \w B^{KL}$. As suggested by Montesinos and Vel\'azquez \cite{MM}, it is natural to consider them both, so we take
\equ
V(B) \equiv -\f16 \left(\f{\lambda}2 \, \eps_{IJKL} + \mu \, \d_{IJKL} \right) B^{IJ} \w B^{KL}.
\nequ

Taking into account the contraint (\ref{epconstraint}), we see that variation by $\phi$  gives rise to the following simplicity constraints,
\equ\label{simplC}
B^{IJ}\w B^{KL} = \f1{12} \eps^{IJKL} \ (B\w \star B).
\nequ
For configurations satisfying the non-degeneracy condition $(B\w \star B) \neq 0$, the above equations
admit two families of solutions, $B^{IJ}=\pm \, e^I\w e^J$ and $B^{IJ}=\pm \star e^I\w e^J$. 

In both cases, the 20 constraints \Ref{simplC} reduce the initial 36 components of $B^{IJ}_{\mu\nu}$ down to 16, parametrized by a tetrad $e^I_\mu$.
The counting shows the need for the extra condition \Ref{epconstraint} on $\phi^{IJKL}$, as otherwise the Lagrange multiplier
would kill too many components of the $B$'s. Without this condition, $\phi$ has 21 components 
and lives in the symmetric tensor product of two adjoint representations of $\mathfrak{so(4)}$. This is given,
using the isomorphism $\mathfrak{so(4)}\cong \mathfrak{su(2)}\oplus \mathfrak{su(2)}$ and the familiar decomposition into irreps of SU(2), by
\equ\label{IJKLdec}
\big[(1,0)\oplus(0,1)\big]\otimes_{\rm Sym} \big[(1,0)\oplus(0,1)\big] =
(2,0)\oplus(0,2)\oplus(1,1)\oplus(0,0)\oplus(0,0).
\nequ
Notice the presence of two singlets in this decomposition: a basis of tensors in the two dimensional
invariant subspace is given by $\d^{IJKL}$ and $\eps^{IJKL}$. Choosing as  above the condition $\eps_{IJKL}\phi^{IJKL}=0$,
we eliminate one of the two singlets. Alternatively, one 
can be more general and, as remarked in \cite{CapBF}, require a linear combination of the singlets to vanish, 
\equ\label{phiCCap}
\left( \d_{IJKL} + a \, \eps_{IJKL} \right) \phi^{IJKL}= 0.
\nequ
The interest of this choice is that it mixes the two families of solutions $e\w e$ and $\star e \w e$. 
Indeed, the constraints now are \cite{CapBF}
\begin{subequations}\label{CCap}\begin{align}
& B^{IJ}\w B^{KL} = \f1{6} \d^{IJKL} \, (B\w B) + \f1{12} \eps^{IJKL} \, (B\w \star B), \label{CCap1} \\ 
& 2a \, (B\w B) = (B\w \star B). \label{CCap2}
\end{align}\end{subequations}
On non-degenerate configurations they are solved by
\equ\label{Bab}
B^{IJ} = (\star + \gamma) e^I \w e^J,
\nequ
with $\ga$ satisfying
\equ\label{a12}
a = \f{1+\ga^2}{4\ga}.
\nequ
This can be easily checked using \Ref{Bab} to compute
\eqa
(B\w B) = 24\, e \, \ga, \qquad
(B\w \star B) = 12 \, e \, (1+\ga^2), \label{BsB}
\neqa
and plugging it in \Ref{CCap2}.

Notice that \Ref{a12} is solved by $\ga = \ga_{\pm} = 2 a \pm \sqrt{4 a^2 - 1}$, $\ga_- \equiv 1/\ga_+$, so
for any choice of $|a|\geq 1/2$ the theory admits two solutions with real $\ga$. This quantity is easily related to the Immirzi parameter inserting the on-shell value of $B$ \Ref{Bab} into the original action \Ref{S}: Using the fact that $\phi_{IJKL} B^{IJ}\w B^{KL}$ vanishes on-shell, we get
\equ\label{S1}
S(e,\om) = \int (\star + \ga) \, e^I\w e^J \w F_{IJ} -2 \, e \left(\la (1+\ga^2) +2\mu\ga \right). \\
\nequ
This is the Einstein-Cartan action with the Holst term \cite{Holst}
$e^I \w e^J \w F_{IJ}$ and cosmological constant term. We can thus identify $\ga$ as (the inverse of) the Immirzi parameter, and the cosmological constant
\equ\label{cc}
\Lambda = \la (1+\ga^2) + 2\mu\ga.
\nequ
Solving the field equation for the spin connection, \Ref{S1} gives
\equ
S(e) = \int e ( R - 2 \Lambda )  + \ga \, \eps \cdot R ,
\nequ
where $\eps \cdot R \equiv \eps^{\mu\nu\rho\sigma} R_{\mu\nu\rho\sigma}$ vanishes thanks to the first Bianchi identities $R_{[\mu\nu\rho]\sigma}=0$. We therefore see that the Plebanski formalism gives back general relativity. 
It is interesting to notice that the cosmological constant $\Lambda$ depends on the Immirzi parameter, and thus on the choice of one of the two solutions for $\ga$ once the free parameters $\la, \mu$ and $a$ are fixed.

The generalization of the Plebanski mechanism to include an Immirzi parameter was the main motivation
for the authors of \cite{CapBF} to consider the ``skewed'' condition \Ref{phiCCap}.
Here we have also included a cosmological constant term, and we notice that the cosmological constant
depends on the Immirzi parameter. More surprises are awaiting if we look at the field equations, and the on-shell value taken by the Lagrange multiplier.

Before continuing, let us make a remark on the limiting choice $a=1/2$. This gives the unique solution $\ga=1$ and corresponds to the self-dual theory \cite{Plebanski,Capo2,Mike}. This is a special case with very different properties, and we do not consider it here. In the following, we will always assume $\ga\neq 1$. 
For a derivation of the field equations in the self-dual case, see e.g. \cite{KrasnovPleb}.

\subsection{Field equations}
Together with \Ref{CCap}, the other field equations are
\eqa
&& d_\om B = 0, \label{Cart} \\
&& F_{IJ}(\om) = \phi_{IJKL} \, B^{KL} +\f13 \left(\f{\lambda}2 \, \eps_{IJKL} + \mu \, \d_{IJKL} \right) B^{KL} \label{EqF1}
\neqa
When $B=(\star + \ga)e \w e$, the Cartan equation \Ref{Cart} is solved by the 
spin connection $\om(e)$, independently of $\ga$.\footnote{The independence of $\ga$ comes from
the $\SO(4)$-invariant nature of $\d^{IJ}_{KL}$ and $\eps^{IJ}_{KL}$.}
Consequently, 
\equ\label{FRiem}
R_{ \mu\nu\rho\sigma}(e) \equiv e_I^{\rho} e_J^{\sigma} F^{IJ}_{\mu\nu}(\om(e))
\nequ 
is the Riemann tensor, and the \Ref{EqF1} contain the Einstein's equations. To see this, it is
convenient to split the 36 equations in \Ref{EqF1} into 20 which will identify the on-shell value of the Lagrange multiplier, and 16 which will give the dynamics for the tetrad $e^I_\mu$. This splitting is achieved
contracting \Ref{EqF1} respectively with $e^I_\rho \, e^J_\sigma$ and
with $\eps^{\mu\nu\rho\sigma} \eps_{IJKL} e^K_\rho$, as we now show in detail.\footnote{An alternative way consists in splitting the equations \Ref{EqF1} into self-dual and antiself-dual components, see e.g. \cite{KrasnovPleb}.}

\subsubsection{First projection}
Contracting $F_{IJ\mu\nu}$ in \Ref{EqF1} with $e^I_\rho \, e^J_\sigma$ we obtain
\equ\label{Riem}
R_{\mu\nu\rho\sigma} = 2 P^{\la\tau}_{\rho\sigma} \, \phi_{\mu\nu\la\tau}
+\f23 (\la + \mu\ga) g_{\mu[\rho} g_{\sigma]\nu} + \f23 (\la\ga+\mu) \f1{2e} \eps_{\mu\nu\rho\sigma},
\nequ
where we defined 
\equ
P^{\la\tau}_{\rho\sigma} \equiv \left(\ga \, \d^{\la\tau}_{\rho\sigma} + \f{1}{2e} \, \eps^{\la\tau}_{\rho\sigma}\right), \qquad 
(P^{-1})^{\la\tau}_{\rho\sigma} \equiv \f1{\ga^2-1}
\left(\ga \, \d^{\la\tau}_{\rho\sigma} - \f{1}{2e} \, \eps^{\la\tau}_{\rho\sigma}\right).
\nequ

From \Ref{Riem} we see that the Bianchi identities for the Riemann tensor imply further field equations on the Lagrange multiplier. In particular,
\equ\label{BI}
0 = \eps^{\mu\nu\rho\la} \, R_{\mu\nu\rho\sigma} = 
4 e \left(\d_{\mu\nu}^{\rho\la} + \f{\ga}{2e} \, \eps^{\rho\la}_{\mu\nu} \right) \phi^{\mu\nu}_{\rho\sigma}
+ 2 e (\la \ga + \mu) \d^\la_\sigma.
\nequ
This means that on top of the constraint \Ref{phiCCap}, the Lagrange multiplier $\phi$ also satisfies the algebraic equation
\equ\label{Eqphi}
\left(\d_{\mu\nu}^{\rho\la} + \f{\ga}{2e} \, \eps^{\rho\la}_{\mu\nu} \right) \phi^{\mu\nu}_{\rho\sigma}
= \f12 (\la \ga + \mu) \d^\la_\sigma.
\nequ
Thanks to the symmetries of $\phi$ recalled above, both index contractions appearing in this equations are proportional
to the identity, namely
\begin{align} \label{Eqphi2}
& \phi^\la_\sigma := \d_{\mu\nu}^{\rho\la} \, \phi^{\mu\nu}_{\rho\sigma} = \f14\phi \, \d^\la_\sigma, 
& (\eps \cdot \phi)^\la_\sigma := \eps^{\rho\la}_{\mu\nu} \, \phi^{\mu\nu}_{\rho\sigma} =\f14 (\eps\cdot\phi) \, \d^\la_\sigma.
\end{align}
The traces can be evaluated solving the system of \Ref{phiCCap} and the trace of \Ref{Eqphi}, giving
\begin{align}\label{AB}
& \phi=-2(1+\ga^2) \f{\la \ga +\mu}{1-\ga^2}, & (\eps\cdot\phi) = 4 \ga \f{\la \ga +\mu}{1-\ga^2}.
\end{align}

Finally, from \Ref{Riem} we also have an expression for the Lagrange multiplier in terms of the Riemann tensor,
\equ\label{phi1}
\phi_{\mu\nu\rho\sigma} = \f12 (P^{-1})^{\la\tau}_{\rho\sigma} R_{\la\tau\mu\nu} 
- \f{\mu}3 g_{\mu[\rho} g_{\sigma]\nu} - \f{\la}{6e}\eps_{\mu\nu\rho\sigma}.
\nequ

\subsubsection{Second projection}

To obtain the Einstein's equations, we now project \Ref{EqF1} along 
$\eps^{\mu\nu\rho\sigma} \eps_{IJKL} e^K_\rho = 6 e e^{[\mu}_I e^\nu_J e^{\sigma]}_L$. 
After dividing by $-4 e$, we find
\eqa
R^\sigma_L - \f12 e^\sigma_L R &=& 2 \ga (\phi^\sigma_L - \f12 \phi \, e^\sigma_L ) 
+ \f{1}{e} (\eps^{\al\be\ga\sigma} \phi_{\al\be\ga L} - \f12 \eps\cdot \phi \, e^\sigma_L) - 
(\la + \mu\ga) e^\sigma_L= \nonumber
\\ &=& \label{EE1}
- \left(\la (1+\ga^2)+2\mu\ga\right) \, e^\sigma_L,
\neqa
where we used \Ref{Eqphi2} and \Ref{AB}. The right hand side can be recognized to be the Einstein equations with cosmological constant given by \Ref{cc}, thus confirming that going on-shell for $B$ in the non-degenerate sector gives general relativity, and in particular the dependence $\Lambda=\Lambda(\la,\mu,\ga)$ found at the level of the action.

Finally, the on-shell value of the Lagrange multiplier is obtained using the Einstein's equations \Ref{EE1} in \Ref{phi1}. This gives
\equ\label{phiOSfinal}
\phi_{\mu\nu\rho\sigma} = \f1{1-\ga^2}\left[\f12\left(-\ga \d^{\la\tau}_{\mu\nu} + \f1{2e}\eps^{\la\tau}_{\mu\nu}\right) C_{\la\tau\rho\sigma} 
- \f{\ga \la + \mu}3  \left((1+\ga^2)g_{\mu[\rho} g_{\sigma]\nu} - \f{\ga}{e}\eps_{\mu\nu\rho\sigma} \right) \right]
\nequ
where $C_{\mu\nu\rho\sigma}$ is the Weyl tensor. On solutions of the field equations, the Lagrange multiplier is the Weyl tensor plus a trace. The latter depends on the two parameters $\la$ and $\mu$ as well as on $\ga$, and can be made to vanish choosing $\ga\la+\mu=0$.

For completeness, let us remind the reader that in the self-dual case, corresponding to $\ga=1$ (or $\ga=i$ if we were using Lorentzian signature), the on-shell value of the Lagrange multiplier is the self-dual part of the Weyl tensor, and remains traceless also if a cosmological constant is present \cite{KrasnovPleb}.

In conclusions, we found that when both the cosmological constant and the Immirzi parameter are present, the Lagrange multiplier acquires an extra trace part. We now discuss the consequences of this for a certain class of modified theories of gravity based on the action \Ref{S}.

\section{Modified Plebanski theory}\label{SecMod}
We now consider a class of modified theories of gravity that can be obtained considering a generic potential $V(\phi, B)$ which depends also on the Lagrange multiplier $\phi$. The same modification structure has been investigated in the self-dual case by Krasnov \cite{Krasnov} (see also \cite{Bengtsson,Freidel,Ishibashi}), and related to previous work by Capovilla, Bengtsson and Peldan \cite{Capo3, Bengtsson-Peldan}. An interesting aspect of the modification is that in the self-dual case it leads to only two propagating degrees of freedom \cite{Krasnov}, like general relativity. On the other hand, the modification in the non-chiral, $\mathfrak{so(4)}$ action \Ref{S} leads to eight degrees of freedom, as showed by the canonical analysis performed in \cite{Alexandrov}.
The additional six degrees of freedom have been interpreted in \cite{noi2} as a second, massive spin two field and a scalar mode.
Therefore the modification of the non-chiral action \Ref{S} to describe gravity leads to stronger departures than in the self-dual case. 

In this paper, we are interested in the modification of \Ref{S} as used in the unification scheme proposed in \cite{Lee}. To that end, we take $V(\phi, B) \equiv \f{g}2 \, {\rm Tr}\phi^2 \, (B \w B)$, where Tr$\phi^2 \equiv \phi^{IJKL} \phi_{IJKL}$ and $g$ is a new coupling constant with dimensions $[\ell^2]$. That is, we consider the action
\equ\label{Smod}
S(B,\om,\phi) = \int B^{IJ} \w F_{IJ} - \f12 \phi_{IJKL} \, B^{IJ} \w B^{KL}
+ \f{g}2 \, {\rm Tr}\phi^2 \, (B \w B).
\nequ
This action was introduced by one of us in \cite{Lee}, where it was shown that with a suitable extension of the gauge group, it could serve for a unification of gravity and Yang-Mills, a line of investigation further pursued in \cite{noi1}.\footnote{The interest in extending the gauge group is the reason not to include the term $(B\w \star B)$, as the latter is peculiar to $\mathfrak{so(4)}$.} 
Here we use the results derived in the previous Section to show that de Sitter spacetime is an exact solution also of the modified theory with $g\neq 0$. This result is used in \cite{Lee,noi1}.

The first thing to remark about \Ref{Smod} is that the variation with respect to $\phi$ does not lead anymore to constraints for $B$, but rather to algebraic relations determining $\phi$ in terms of $B$. As such, there is no more rationale for imposing additional restrictions on $\phi$, such as (\ref{epconstraint}) or (\ref{phiCCap}). 
This is a key difference with the usual Plebanski formulation and we see this as a nicety of the extended theory, because it reduces the conditions to be imposed by hand on the fundamental theory.  

As we show below, even dropping (\ref{phiCCap}), we can still introduce the Immirzi parameter through a choice of
an ansatz to the field equations for $\phi^{IJKL}$. 
However, it turns out that there is a further interesting consequence.  Dropping (\ref{phiCCap}) means that 
$\phi^{IJKL}$ now has twenty-one components
rather than twenty, so that there is one more field equation than before.  We will see that a consequence of
this additional field equation is that, when the cosmological constant is non-vanishing, the Immirzi parameter is 
constrained to satisfy a relation, which has just a few solutions.  

It was partly to understand how this arises that we took the time to straighten out the interplay of the 
cosmological constant and Immirzi parameter in the original non-chiral Plebanski theory.

\subsection{DeSitter solution}

Taking the variation of \Ref{Smod} with respect to an  unconstrained $\phi$, we get 
\equ\label{ModEq1bis}
 B^{IJ}\w B^{KL} = 2\, g \, \phi^{IJKL} \, (B\w B).  
\nequ
The $\om$ variation is unchanged, and the $B$ variation gives
\equ
F_{IJ} = \phi_{IJKL} B^{KL} - g \, {\rm Tr} \phi^2 \, B_{IJ}. \label{ModEq4}
\nequ
These are the field equations of the modified theory of gravity, and it is to be expected that solutions will differ from the ones of general relativity. However we now show that DeSitter space is an exact solution of the theory.\footnote{The reader familiar with the modification in the self-dual theory might expect such a result, on the basis that Weyl vanishes on DeSitter space. However the situation turns out to be more subtle in our case, due to the presence of the Immirzi parameter.}

To study this system, we take the following ansatz as a solution of the full field equations,
\equ\label{Lee}
\phi^{IJKL} = \f1{12g} \left( \d^{IJKL}+ \f{1+\ga^2}{4\ga}\eps^{IJKL}\right).
\nequ
Plugging this ansatz into \Ref{ModEq1bis}, we obtain exactly \Ref{CCap}, which we already know is solved by \Ref{Bab} in the non-degenerate sector. Restricting to this sector, we can solve \Ref{Cart} as before and find that $F(\om(e))$ gives the Riemann tensor. We can now attempt to solve \Ref{ModEq4} and check whether \Ref{Lee} was a good ansatz.

To do so, we first evaluate
\equ
{\rm Tr} \phi^2 = \f1{4! g^2} \, \f{\ga^4+6\ga^2+1}{4\ga^2}.
\nequ
This result together with \Ref{Bab} and \Ref{Lee} allows us to rewrite \Ref{ModEq4} as
\equ\label{EqMot2}
F^{IJ}_{\mu\nu}(\om(e)) = \f{1}{4! g} \f{1}{2\ga} \left[ f_1(\ga) \d^{IJ}_{KL} + f_2(\ga) \f12\eps^{IJ}_{KL} \right] 
e^K_\mu e^L_\nu,
\nequ
where
\equ
f_1(\ga) = -\ga^4+6\ga^2+3, \qquad f_2(\ga) = \f1{\ga} (3\ga^4+6\ga^2-1).
\nequ
If we now project along $e_\rho^I e_\sigma^J$ and use \Ref{FRiem} we get
\equ\label{Riem2}
R_{\mu\nu\rho\sigma}(g_{\mu\nu}) =  \f{1}{4! g} \f{1}{2\ga} 
\left[ f_1(\ga) g_{\mu[\rho} g_{\sigma]\nu} + f_2(\ga) \f1{2e} \eps_{\mu\nu\rho\sigma} \right].
\nequ
These field equations manifestly violate the first Bianchi identities, unless we fix $\ga=\ga_0$ such that $f_2(\ga_0)=0$.
This equation admits two real roots given by $\ga_0 = \pm (2/{\sqrt{3}}-1)^{1/2}\simeq \pm 0.39$. 
For this value, $f_1(\ga_0)= 16 (\sqrt{3}-1)/3$.
Tracing \Ref{Riem2} we then get
\equ\label{Ricci}
R_{\mu\nu}(g_{\mu\nu}) =  \f{\sqrt{3}-1}{6 g \ga_0} g_{\mu\nu},
\nequ
which are the Einstein equations with cosmological constant equal to
$
\Lambda = ({\sqrt{3}-1})/{6 g \ga_0}. 
$

Next, we see from \Ref{Riem2} that the Weyl tensor vanishes. From this fact and \Ref{Ricci},
we conclude that DeSitter spacetime is an exact solution of the modified theory, with the Lagrange multiplier taking the value \Ref{Lee}.

It goes without saying that the modified theory will have other solutions with non-vanishing Weyl tensor, but they will give a different value of $\phi$ on-shell, and more importantly, they will differ from solutions of general relativity. 


In conclusions, we can describe DeSitter/anti-DeSitter, but this requires fixing the Immirzi parameter $\ga$ in the ansatz \Ref{Lee} to the value $\ga_0\simeq \pm 0.39$, and thus also in the solution \Ref{Bab} for $B$. The resulting cosmological constant is nonetheless still free to vary according to $g$, as well as the choice of sign of $\ga_0$.

\subsection{Reintroducing the Immirzi parameter}

Before concluding, it is interesting to note that extra terms can be added to the action,
which allow one to free the Immirzi parameter. 
In particular, the extra kinetic term
\equ\label{Imm2}
\ga_1 \, \eps_{IJKL} B^{IJ} \w F^{KL},
\nequ
as well as the alternative volume form in the potential,
\equ\label{Imm3}
\ga_2 \, {\rm Tr}\phi^2 \, (B \w \star B).
\nequ
Adding one or both of these terms results in a dependence on the extra parameters $\ga_1$ and/or $\ga_2$ of the function $f_2$ in \Ref{EqMot2}. This allows one to find a family of real solutions $\ga_i(\ga)$ for which $f_2\equiv 0$ without fixing $\ga$ to a finite range of values.
In this way, the fixation of the Immirzi parameter can be avoided.

We should however point out that this possibility is not available in view of the unified theory of \cite{Lee,noi1}, because it relies on the existence of a second singlet in \Ref{IJKLdec}, the completeley antisymmetric tensor $\eps^{IJKL}$. The presence of two singlets in the decomposition \Ref{IJKLdec} is a peculiarity of $\mathfrak{so(4)}$, which for instance does not generalize $\mathfrak{so(n)}$ with $n=8$ or $n=10$, which are cases of interest for \cite{noi1}. Therefore terms like \Ref{Imm2} or \Ref{Imm3} would break explicitly the gauge invariance of the action.

\section{Conclusions}
We studied the field equations of the non-chiral Plebanski action for general relativity in the presence of a cosmological constant and an Immirzi parameter. The latter is introduced using the idea proposed in \cite{CapBF} of removing the undesired 21st component of the Lagrange multiplier via a linear combination of the two possible singlets, $\d_{IJKL}$ and $\eps_{IJKL}$. 
We find that the cosmological constant depends on the values of the free parameters in the initial action as well as on the Immirzi parameter. We also find that the Lagrange multiplier takes on-shell a value given by a self and antiself-dual combination of the Weyl tensor, plus a trace term. The novelty is the presence of the trace in the Lagrange multiplier, and the fact that both the trace and the cosmological constant depend on the Immirzi parameter.
The relevant formulas are given by \Ref{EE1} and \Ref{phiOSfinal}, and to the best of our knowledge had never appeared in the literature.

Then, we discussed some implications of this result for a class of modified theories of gravity introduced in \cite{Lee}. In particular, we showed that DeSitter spacetime is an exact solution for the modified theory if we fix the Immirzi parameter to a specific value. The cosmological constant is still allowed to freely vary by varying the new coupling constant introduced by the modification of the action.
The result is used in \cite{Lee, noi1} to support the extended Plebanski action as a possible action for a unification of gravity and Yang-Mills theory.

\bigskip


{\bf Acknowledgements}. We are grateful to Garrett Lisi for collaboration on a project out of which these results emerge and we thank him for his
insightful comments.  Research at Perimeter Institute for Theoretical 
Physics is supported in part by the Government of Canada through NSERC and by the 
Province of Ontario through MRI.



\begin{thebibliography}{99}

\bibitem{Deser}
 S.~Deser,
 ``Self-interaction and gauge invariance,''
 Gen.\ Rel.\ Grav.\  {\bf 1}, 9 (1970)
 [arXiv:gr-qc/0411023].
 
\bibitem{Plebanski}  
M. J. Pleba\'{n}ski, ``On the separation of Einsteinian substructures,''
J. Math. Phys. {\bf 18} (1977), 2511.

\bibitem{Capo2}
  R.~Capovilla, T.~Jacobson, J.~Dell and L.~Mason,
  ``Selfdual two forms and gravity,''
  Class.\ Quant.\ Grav.\  {\bf 8} (1991) 41.

\bibitem{Krasnov}
  K.~Krasnov,
  ``Renormalizable Non-Metric Quantum Gravity?,''
  arXiv:hep-th/0611182;
\
  K.~Krasnov,
  ``Non-Metric Gravity I: Field Equations,''
  Class.\ Quant.\ Grav.\  {\bf 25} (2008) 025001
  [arXiv:gr-qc/0703002];
\  K.~Krasnov and Y.~Shtanov,
  ``Non-Metric Gravity II: Spherically Symmetric Solution, Missing Mass and Redshifts of Quasars,''
  Class.\ Quant.\ Grav.\  {\bf 25} (2008) 025002
  [arXiv:0705.2047 [gr-qc]];
\  
K.~Krasnov,
  ``On deformations of Ashtekar's constraint algebra,''
  Phys.\ Rev.\ Lett.\  {\bf 100} (2008) 081102
  [arXiv:0711.0090 [gr-qc]];
\
  K.~Krasnov,
  ``Plebanski gravity without the simplicity constraints,''
  Class.\ Quant.\ Grav.\  {\bf 26} (2009) 055002
  [arXiv:0811.3147 [gr-qc]];
\  K.~Krasnov,
  ``Motion of a 'small body' in non-metric gravity,''
  arXiv:0812.3603 [gr-qc].

\bibitem{Mike}
  M.~P.~Reisenberger,
  ``New Constraints For Canonical General Relativity,''
  Nucl.\ Phys.\  B {\bf 457}, 643 (1995)
  [arXiv:gr-qc/9505044];
\
  M.~P.~Reisenberger,
  ``Classical Euclidean general relativity from *left-handed area =
  right-handed area*,''
  arXiv:gr-qc/9804061.

\bibitem{De Pietri}
  R.~De Pietri and L.~Freidel,
  ``so(4) Pleba\'{n}ski Action and Relativistic Spin Foam Model,''
  Class.\ Quant.\ Grav.\  {\bf 16} (1999) 2187
  [arXiv:gr-qc/9804071].

\bibitem{Lee}
  L.~Smolin,
  ``The Plebanski action extended to a unification of gravity and Yang-Mills theory,''
  arXiv:0712.0977 [hep-th].

\bibitem{Alexandrov}
  S.~Alexandrov and K.~Krasnov,
  ``Hamiltonian Analysis of non-chiral Plebanski Theory and its Generalizations,''
  Class.\ Quant.\ Grav.\  {\bf 26} (2009) 055005
  [arXiv:0809.4763 [gr-qc]].
  
\bibitem{noi2}
L. Smolin and S. Speziale, ``Bi-metric theory of gravity from the non-chiral Plebanski action,'' to appear.

\bibitem{noi1}
A. Garrett Lisi, L. Smolin and S. Speziale, ``Unification of gravity, gauge interactions and Higgs bosons in the extended Plebanski formalism,'' to appear.

\bibitem{MM}
M. Montesinos and M. Vel\`azquez, ``BF gravity with Immirzi parameter and
cosmological constant,'' to appear.

\bibitem{CapBF}
  R.~Capovilla, M.~Montesinos, V.~A.~Prieto and E.~Rojas,
  ``BF gravity and the Immirzi parameter,''
  Class.\ Quant.\ Grav.\  {\bf 18} (2001) L49
  [Erratum-ibid.\  {\bf 18} (2001) 1157]
  [arXiv:gr-qc/0102073].

\bibitem{Holst}
  S.~Holst,
  ``Barbero's Hamiltonian derived from a generalized Hilbert-Palatini action,''
  Phys.\ Rev.\  D {\bf 53}, 5966 (1996)
  [arXiv:gr-qc/9511026].

\bibitem{KrasnovPleb}
  K.~Krasnov,
  ``Pleba\'{n}ski Formulation of General Relativity: A Practical Introduction,''
  arXiv:0904.0423 [gr-qc].
  
\bibitem{Bengtsson}
  I.~Bengtsson,
  ``Selfduality and the metric in a family of neighbors of Einstein's equations,''
  J.\ Math.\ Phys.\  {\bf 32}, 3158 (1991);
\ 
  I.~Bengtsson,
  ``Note on non-metric gravity,''
  Mod.\ Phys.\ Lett.\  A {\bf 22} (2007) 1643
  [arXiv:gr-qc/0703114].

\bibitem{Freidel}
  L.~Freidel,
  ``Modified gravity without new degrees of freedom,''
  arXiv:0812.3200 [gr-qc].

\bibitem{Ishibashi}
  A.~Ishibashi and S.~Speziale,
  ``Spherically symmetric black holes in minimally modified self-dual gravity,''
  Class.\ Quant.\ Grav.\  {\bf 26}, 175005 (2009)
  [arXiv:0904.3914 [gr-qc]].

\bibitem{Capo3}
  R.~Capovilla,
  ``Generally Covariant Gauge Theories,''
  Nucl.\ Phys.\  B {\bf 373}, 233 (1992).

\bibitem{Bengtsson-Peldan}
  I.~Bengtsson,
  Max Born Symposium 1992, pp. 183-188 [arXiv:gr-qc/9210001];
\
  P.~Peldan,
  ``Actions for gravity, with generalizations: A Review,''
  Class.\ Quant.\ Grav.\  {\bf 11} (1994) 1087
  [arXiv:gr-qc/9305011].


\end{thebibliography}
\end{document}